# Shortest Path and Distance Queries on Road Networks: An Experimental Evaluation

Lingkun Wu[†], Xiaokui Xiao[†], Dingxiong Deng[§], Gao Cong[†], Andy Diwen Zhu[†],
Shuigeng Zhou[§]

[†]School of Computer Engineering
Nanyang Technological University
Singapore

[§]School of Computer Science
Fudan University
China

[†]{wulingkun, xkxiao, gaocong, dwzhu}@ntu.edu.sg   [§]{dxdeng, sgzhou}@fudan.edu.cn

## ABSTRACT

Computing the shortest path between two given locations in a road network is an important problem that finds applications in various map services and commercial navigation products. The state-of-the-art solutions for the problem can be divided into two categories: *spatial-coherence-based* methods and *vertex-importance-based* approaches. The two categories of techniques, however, have not been compared systematically under the same experimental framework, as they were developed from two independent lines of research that do not refer to each other. This renders it difficult for a practitioner to decide which technique should be adopted for a specific application. Furthermore, the experimental evaluation of the existing techniques, as presented in previous work, falls short in several aspects. Some methods were tested only on small road networks with up to one hundred thousand vertices; some approaches were evaluated using *distance queries* (instead of shortest path queries), namely, queries that ask only for the length of the shortest path; a state-of-the-art technique was examined based on a faulty implementation that led to incorrect query results.

To address the above issues, this paper presents a comprehensive comparison of the most advanced spatial-coherence-based and vertex-importance-based approaches. Using a variety of real road networks with up to twenty million vertices, we evaluated each technique in terms of its preprocessing time, space consumption, and query efficiency (for both shortest path and distance queries). Our experimental results reveal the characteristics of different techniques, based on which we provide guidelines on selecting appropriate methods for various scenarios.

## 1. INTRODUCTION

Computing the shortest path between two locations in a road network is an important problem that finds applications in various map services and commercial navigation products. The classic solution for the problem is Dijkstra's algorithm [9], which, given a source $s$ and a destination $t$ in a road network $G$, traverses the vertices in $G$ in ascending order of their distances to $s$. Once $t$ is reached during the traversal, the shortest path from $s$ to $t$ is computed and returned. This algorithm is simple and elegant, but it is often inefficient for sizeable road networks [21]. The reason is that, when computing the shortest path from $s$ to $t$, Dijkstra's algorithm has to visit all vertices in $G$ that are closer to $s$ than $t$, and the number of such vertices can be enormous when $s$ and $t$ are far apart.

Over the past two decades, a plethora of techniques have been proposed to address the deficiency of Dijkstra's algorithm by exploiting the characteristics (*e.g.*, planarity) of road networks [4–8, 10–26]. In particular, the state-of-the-art approaches can be classified into two categories. Algorithms in the first category [21, 23–25] take advantage of the fact that shortest paths in road networks are often *spatially coherent*. To illustrate the concept of *spatial coherence*, let us consider four locations $s$, $s'$, $t$, and $t'$ in a road network. If $s$ is close to $s'$ and $t$ is close to $t'$, then the shortest path from $s$ to $t$ is likely to share vertices with the shortest path from $s'$ to $t'$. Such spatial coherence of shortest paths makes it possible to compress all shortest paths in a road network in a concise format, and the compressed paths can be used to answer queries efficiently. Representative spatial-coherence-based algorithms include *Spatially Induced Linkage Cognizance (SILC)* [21, 23] and *Path-Coherent Pairs Decomposition (PCPD)* [25].

Methods in the second category [5, 6, 11, 20, 22, 26], on the other hand, are built upon the observation that certain vertices in a road network are more important for shortest path queries. For example, a vertex that represents the entrance of a highway tends to be accessed much more frequently (in shortest path queries) than a vertex that corresponds to a road junction in a countryside. This observation motivates various approaches [5–7, 11, 20, 22, 26] that (i) order the vertices in a road network in terms of their importance, and (ii) pre-compute the shortest paths among the important vertices to accelerate query processing. Among those approaches, *Contraction Hierarchies (CH)* [11] and *Transit Node Routing (TNR)* [5, 6, 26] are shown to be the most efficient.

**Motivations.** Spatial-coherence-based algorithms and vertex-importance-based methods were both demonstrated to significantly outperform Dijkstra's algorithm [5–7, 11, 20–26]. To our surprise, however, the two categories of techniques have not been compared systematically under the same experimental platform. This seems to be caused by the fact that the two categories of methods were developed from two independent lines of research that do not refer to each other. As a consequence, the relative superior of two kinds of approaches remain unclear, which renders it difficult for a practitioner to decide which technique should be adopted for a specific application. Furthermore, there exist several other reasons that motivate a more thorough evaluation of each technique.

First, the performance of the state-of-the-art spatial-coherence-based algorithms, SILC [21, 23] and PCPD [25], were tested using only small road networks with up to one hundred thousand vertices. Therefore, it remains open whether the algorithms can scale to large road networks (with millions of vertices) commonly used in modern map applications. Second, empirical studies on vertex-importance-based methods [5–7,11,20,22,26] largely focus on *distance queries*, which concern about the length of the shortest path between two given locations, instead of the sequence of the edges that comprises the shortest path. Consequently, there is a need for an extensive assessment of the efficiency of vertex-importance-based methods for shortest path queries. Third, TNR [5, 6], a state-of-the-art vertex-importance-based method, adopts a faulty preprocessing algorithm that leads to incorrect answers for shortest path and distance queries (see Appendix B for a discussion). This invalidates the experimental results previously reported for TNR [5], and motivates a re-examination of the technique. All of the aforementioned issues call for a more comprehensive evaluation of the existing techniques for shortest path and distance queries.

**Contributions.** This paper presents an experimental comparison of the state-of-the-art spatial-coherence-based algorithms (*i.e.*, SILC [21, 23] and PCPD [25]) and vertex-importance-based methods (*i.e.*, CH [11] and TNR [5]). Using a variety of real road networks with up to twenty million vertices, we evaluated the performance of each technique in terms of its preprocessing time, space consumption, and query efficiency (for both shortest path and distance queries). Our experimental results reveal the characteristics of different techniques, based on which we provide guidelines on selecting appropriate methods for various scenarios. In addition, we analyzed the defect of the preprocessing algorithm adopted by TNR, and we proposed a correction of the algorithm that works well in our experiments.

The remainder of the paper is organized as follows. Section 2 introduces the concepts and notations frequently used in the paper. Section 3 reviews the techniques that we evaluated. Section 4 reports the experimental results. Section 5 concludes this paper. The appendix covers additional related work and experimental results, as well as an analysis of the defect of TNR, along with the details of our implementations.

## 2. PROBLEM DEFINITION

Let $G$ be a road network (*i.e.*, a degree-bounded connected graph) with an edge set $E$ and a vertex set $V$ that contains $n$ vertices. Let each edge $e \in E$ be associated with a weight $w(e)$, which we assume (without loss of generality) to be the length of $e$. For ease of exposition, we consider undirected graphs in this paper.

We study two types of queries on $G$, namely, *shortest path queries* and *distance queries*. Given two vertices $s, t \in V$, a shortest path query asks for a sequence of edges $(e_1, e_2, \ldots, e_k)$ that connects $s$ to $t$, such that $\sum_{i=1}^{k} w(e_i)$ is minimized. On the other hand, a distance query between $s$ and $t$ requests only the value $\sum_{i=1}^{k} w(e_i)$, such that $e_1, e_2, \ldots, e_k \in E$ comprise the shortest path from $s$ to $t$. Distance queries have been the focus of previous work [5–7, 11, 22, 26], and it is useful in the scenario where the distance between two locations instead of the shortest path is the major concern. For example, assume that a user has a list of her favorite Italian restaurants, and she wants to identify the restaurant that is closest to her working place $q$. In that case, she may issue a distance query from $q$ to each of the restaurants to find the nearest one. For convenience, we use $dist(v_1, v_2)$ to denote the length of the shortest path between two vertices $v_1, v_2 \in V$.

## 3. ALGORITHMS

This section reviews the five techniques evaluated in our experiments, namely, (i) the bidirectional Dijkstra's algorithm, a variation of Dijkstra's algorithm that we used as the baseline, (ii) CH [11] and TNR [5], two state-of-the-art vertex-importance-based methods, and (iii) SILC [21, 23] and PCPD [25], two most advanced spatial-coherence-based algorithms. Interested readers are referred to Appendix A for a summary of other existing work on shortest path and distance queries.

### 3.1 Bidirectional Dijkstra's Algorithm

Given two vertices $s, t \in V$, the bidirectional Dijkstra's algorithm [19] invokes two instances of Dijkstra's algorithm simultaneously, such that the first (resp. the second) instance traverses the vertices in $G$ in ascending order of their distances to $s$ (resp. $t$). In addition, the algorithm maintains a minimum spanning tree rooted at $s$ (resp. $t$) for the vertices visited during the first (resp. second) traversal. The two traversals terminate when they meet at a vertex $u \in V$. Let $V_1$ (resp. $V_2$) be the set of vertices visited by the traversal that starts from $s$ (resp. $t$). It can be verified that the shortest path between $s$ and $t$ must either pass through $u$, or go across two adjacent vertices $v_1 \in V_1$ and $v_2 \in V_2$. Therefore, $dist(s, t)$ should equal the smallest value among $dist(s, u) + dist(u, t)$ and $dist(s, v_1) + dist(v_1, v_2) + dist(v_2, v)$, for any two adjacent vertices $v_1 \in V_1$ and $v_2 \in V_2$. Once $dist(s, t)$ is decided, the shortest path between $s$ and $t$ can be retrieved from the spanning trees constructed during the traversals from $s$ and $t$.

As with Dijkstra's algorithm, the bidirectional variant runs in $O(n \log n)$ time (given that $G$ is a degree-bounded connected graph), but it is usually more efficient than Dijkstra's algorithm in practice. This is because, intuitively, each of the two graph traversals invoked by the bidirectional algorithm visit the vertices that are within roughly $dist(s, t)/2$ distance to $s$ or $t$. The number of such vertices is often smaller than the number of vertices that are within $dist(s, t)$ distance to $s$, *i.e.*, the vertices that need to be traversed by Dijkstra's algorithm.

### 3.2 Contraction Hierarchies

Contraction Hierarchies (CH) [11] is a graph indexing technique that imposes a total order on the vertices in $G$ according to their relative importance. It pre-computes the distances among various vertices based on the total order, and it utilizes the pre-computed distances to accelerate shortest path and distance queries. To explain the preprocessing step of CH, let us consider the road network $G$ in Figure 1 that contains eight vertices $v_1, v_2, \ldots, v_8$ and nine edges. In particular, the lengths of the edges $(v_2, v_8)$ and $(v_6, v_8)$ equal 2, while the lengths of the other edges are 1.

Without loss of generality, assume that CH imposes a total order $v_1 < v_2 < \ldots < v_8$ on the vertices in $G$. The preprocessing step of CH examines the vertices following the total order. For each vertex $v_i$, CH first inspects the neighbors of $v_i$ (*i.e.*, the vertices adjacent to $v_i$ in $G$), and checks whether there exist two neighbors $v_j$ and $v_k$, such that the shortest path from $v_j$ to $v_k$ passes through $v_i$. For any such $v_j$ and $v_k$, CH inserts in $G$ an artificial edge $c$ (referred to as a *shortcut*) that connects $v_j$ to $v_k$, such that $w(c) = dist(v_j, v_k)$. The shortcut is tagged with $v_i$ to indicate that it is created when $v_i$ is processed. (The tags of shortcuts are crucial for shortest path queries, as will be clarified shortly.) Notice that, with the shortcut added, $v_j$ becomes a neighbor of $v_k$, and vice versa. Once all neighbors of $v_i$ are examined, $v_i$ is removed from $G$. This process is referred to as the *contraction* of $v_i$ [11]. After all vertices are contracted, CH terminates the preprocessing step, and the shortcuts that have been created during the contraction process are added to the original road network.

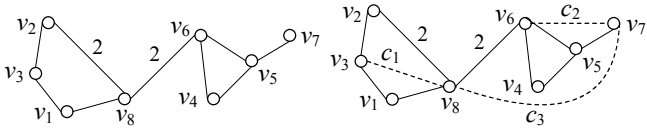

**Figure 1: Road Network**    **Figure 2: Contraction Hierarchies**

For example, given the road network in Figure 1, CH first inspects the vertex $v_1$. $v_1$ has only two neighbors $v_3$ and $v_8$, and the shortest path between $v_3$ and $v_8$ goes by $v_1$. Therefore, when contracting $v_1$, CH connects $v_3$ and $v_8$ with a shortcut $c_1$, such that $w(c_1) = dist(v_3, v_8) = 2$, as illustrated in Figure 2. After $v_1$ is removed from $G$, CH proceeds to examine $v_2$. Notice that $v_2$ has only two neighbors $v_3$ and $v_8$. Given that $v_1$ has been deleted, the shortest path between $v_3$ and $v_8$ consists of the newly constructed shortcut $c_1$, which does not pass through $v_2$. Hence, $v_2$ is removed without introducing any new shortcut into $G$. After that, the contraction process is performed on the other vertices in turn, which leads to two additional shortcuts. In particular, the contraction of $v_5$ (resp. $v_6$) results in a shortcut $c_2$ (resp. $c_3$) that connects $v_7$ to $v_6$ (resp. $v_8$), and $w(c_2) = 2$ (resp. $w(c_3) = 4$). Figure 2 illustrates the road network produced by the preprocessing step of CH.

Given the road network augmented with shortcuts, CH answers a distance query between any two vertices $s, t \in V$ using the bidirectional Dijkstra's algorithm with some minor modifications. In particular, when traversing the vertices in $G$ in ascending order of their distances to $s$ (or $t$), CH considers only those edges and shortcuts that connect a visited vertex $v$ to an unvisited vertex $v'$ whose rank is higher than $v$, i.e., $v < v'$. For example, if we are to find the distance between $v_3$ and $v_7$ in Figure 2, the traversal starting from $v_3$ will visit $v_8$ but not $v_1$ or $v_2$, since $v_1 < v_2 < v_3 < v_8$. Similarly, the traversal from $v_7$ will visit $v_8$ but not any other vertex. After the two traversals meet at $v_8$, both of them terminate since there does not exist an edge or shortcut that connects a visited vertex (i.e., $v_3$, $v_7$, or $v_8$) to an unvisited vertex with a higher rank. Therefore, CH returns the distance between $v_3$ and $v_7$ as $dist(v_3, v_8) + dist(v_7, v_8) = w(c_1) + w(c_3) = 6$. In general, the two traversals may not stop immediately after they meet at a vertex; there exist a few conditions that a traversal should fulfill before it can terminate (see [11] for details). But intuitively, CH is more efficient than the bidirectional Dijkstra's algorithm, as it avoids visiting the vertices with lower ranks in the total order.

The aforementioned algorithm can also be used to compute the shortest path from $s$ to $t$ in the *augmented* road network. The resulting path, however, may contain several shortcuts, and hence, it cannot be returned to the user unless it is transformed to a normal path in $G$. For this purpose, CH examines the tag associated with each shortcut in the path. For a shortcut $c$ that connects two vertices $v_j$ and $v_k$, if its tag indicates that $c$ is created during the contraction of $v_i$, then CH removes $c$ from the path, and replaces it with two edges $(v_j, v_i)$ and $(v_i, v_k)$. For example, the shortcut $c_1$ in Figure 2 is adjacent to two vertices $v_3$ and $v_8$, and it is constructed when $v_1$ is contracted. Therefore, if $c_1$ appears in the shortest path, CH substitutes it with two edges $(v_3, v_1)$ and $(v_1, v_8)$. Since both $(v_3, v_1)$ and $(v_1, v_8)$ are edges in the original graph, the transformation of $c_1$ is accomplished. In general, if $(v_j, v_i)$ (resp. $(v_i, v_k)$) is not an edge in $E$, then it can be verified that $(v_j, v_i)$ (resp. $(v_i, v_k)$) must be a shortcut that has been constructed in the preprocessing step. In that case, CH will recursively replace the shortcut with smaller segments in a similar manner. When all edges remaining in the shortest path are edges in $E$, CH returns the path as the final result.

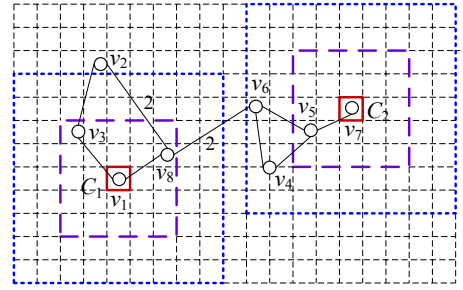

**Figure 3: Transit Node Routing**

Note that the efficiency of CH is determined by the total order on the vertices. An inferior ordering can lead to $O(n^2)$ shortcuts, which in turn results in an $O(n^2 log n)$ time complexity for shortest path and distance queries. Existing work [11] on CH has suggested several heuristic approaches for deriving a favorable ordering based on the distribution of the vertices and edges in $G$.

### 3.3 Transit Node Routing

Transit Node Routing (TNR) [5] is an indexing method that imposes a grid on the road network. It pre-computes the shortest paths from within each grid cell $C$ to a set of vertices that are deemed important for $C$ (those vertices are referred to as the *access nodes* for $C$). In what follows, we elaborate TNR using the example in Figure 3, which shows a grid imposed on the road network in Figure 1.

For each cell $C$ in the grid, let us define the *inner shell* (resp. *outer shell*) of $C$ as the boundary of the $5 \times 5$ (resp. $9 \times 9$) square centered at $C$. For instance, the dashed-line (resp. dotted-line) square on the left of Figure 3 illustrates the inner (resp. outer) shells of the cell that contains $v_1$. A set $A$ of vertices in $V$ is a set of *access nodes* for a grid cell $C$, if and only if it satisfies the following conditions. First, each vertex in $A$ is an endpoint of an edge that intersects the inner shell of $C$. Second, for any shortest path from a vertex in $C$ to a vertex that lies beyond the outer shell of $C$, the path must pass through at least one vertex in $A$, i.e., the vertices in $A$ "cover" all shortest paths from the interior of $C$ to the exterior of its outer shell. For example, in Figure 3, $\{v_3, v_8\}$ (resp. $\{v_5\}$) is a set of access nodes for the cell $C_1$ (resp. $C_2$).

Given the access nodes of all grid cells, TNR pre-computes two sets of distance information: (i) the distance from each vertex $v$ to each access node of the cell that contains $v$, and (ii) the distance between any two access nodes of any two different cells. For instance, given the grid in Figure 3, TNR pre-computes the distances from $v_1$ (resp. $v_7$) to the access nodes of $C_1$ (resp. $C_2$), namely, $dist(v_1, v_3)$, $dist(v_1, v_8)$, and $dist(v_7, v_5)$. In addition, TNR also computes the pairwise distances among the access nodes of $C_1$ and $C_2$, i.e., $dist(v_3, v_5)$ and $dist(v_8, v_5)$.

With the pre-computed distances, TNR can efficiently derive the distance between any two vertices $s, t \in V$, as long as $t$ lies beyond the outer shell of the cell that contains $s$. For example, suppose that we are to compute the distance between $v_1$ and $v_7$ in Figure 3. Since $v_1$ is contained in the cell $C_1$, and since $v_7$ lies in the exterior of $C_1$'s outer shell, the shortest path from $v_1$ to $v_7$ must pass through an access node of $C_1$, i.e., the path must go by either $v_3$ or $v_8$. By the same rationale, the shortest path should also pass through $v_5$, which is the only access node of the cell $C_2$ that encloses $v_7$. Therefore, the distance between $v_1$ and $v_7$ should equal the smaller one of $dist(v_1, v_3) + dist(v_3, v_5) + dist(v_5, v_1)$ and $dist(v_1, v_8) + dist(v_8, v_5) + dist(v_5, v_1)$, both of which can be derived using the pre-computed distances.

In general, given any two cells $C_s$ and $C_t$ such that they are not contained in each other's outer shells, the distance between any vertex $s$ in $C_s$ and any vertex $t$ in $C_t$ can be computed as

$$dist(s,t) = \min_{v_s \in A_s, v_t \in A_t} dist(s,v_s) + dist(v_s,v_t) + dist(v_t,t), \quad (1)$$

where $A_s$ and $A_t$ denote the sets of access nodes for $C_s$ and $C_t$, respectively. On the other hand, if $C_t$ lies inside the outer shell of $C_s$, then TNR cannot derive $dist(s,t)$ based on the pre-computed distances. In that case, we need to resort to other techniques (*e.g.*, CH or the bidirectional Dijkstra's algorithm) to compute $dist(s,t)$.

Interestingly, the aforementioned algorithm (for distance queries) can also be adopted to compute the shortest path from $s$ to $t$. Specifically, we first identify the neighbor $v$ of $s$ that minimizes $dist(s,v) + dist(v,t)$, where $dist(v,t)$ is derived by Equation 1. It can be verified that $v$ should lie on the shortest path from $s$ to $t$. After that, we examine the neighbors of $v$, and pinpoint the neighbor $v'$ that minimizes $dist(v,v') + dist(v',t)$, so on and so forth. With this traversal approach, we can efficiently compute the part of the shortest path that lies outside the outer shell of $C_t$. After that, we can start a similar traversal from $t$ to derive the remaining part of the path. In general, TNR can derive the shortest path between $s$ and $t$ using the pre-computed distances, as long as the outer shells of $C_s$ and $C_t$ do not intersect. Otherwise, an alternative method is required for computing the shortest path.

The performance of TNR depends highly on the granularity of the grid imposed on the road network. A finer grid leads to higher space overhead (due to the increased total number of access nodes), but it also allows more shortest path and distance queries to be answered efficiently using the pre-computed information. (See Appendix E.1 for an experimental evaluation on the effects of grid granularity on the performance of TNR.)

**Remarks.** In our experiments, we adopt the following approach for computing the access nodes for each grid cell $C$. First, we retrieve the edges in $E$ intersecting $C$'s outer shell. Let $V_{out}$ be a set that contains the endpoints of those edges. Then, we compute the shortest path from each vertex in $C$ to each vertex in $V_{out}$. For each shortest path, we identify the edge $e$ on the path that intersects the inner shell of $C$, and we choose an endpoint of $e$ as an access node for $C$. Let $A$ be the set of access nodes that we collect after all shortest paths are examined. It can be verified that any shortest path from within $C$ to beyond the outer shell of $C$ must pass through at least one vertex in $A$, *i.e.*, $A$ is a complete set of access nodes for $C$. Bast et al. [5] propose a method that improves the aforementioned approach in terms of computation time. The improved method, however, is flawed and may lead to incorrect query results (see Appendix B for a discussion).

### 3.4 Spatially Induced Linkage Cognizance

Spatially Induced Linkage Cognizance (SILC) [21, 23] is a technique that (i) pre-computes the all-pairs shortest paths in the road network, and (ii) stores the shortest paths in a concise form for efficient query processing. Consider for example the road network $G$ in Figure 1. Given the shortest paths between all pairs of vertices in $G$, SILC first inspects those shortest paths that share the same starting point. For example, let us consider the paths from the vertex $v_8$ in Figure 1 to the other vertices. Observe that each of the shortest paths goes by a neighbor of $v_8$. In particular, the paths from $v_8$ to $v_4, v_5, v_6, v_7$ pass through $v_6$, while the paths from $v_8$ to $v_1$ and $v_3$ are via $v_1$. Let us partition the vertices in $V \setminus \{v_8\}$ into equivalence classes, such that for any vertex in the same equivalence class, its shortest path to $v_8$ passes through the same neighbor of $v_8$. Figure 4 illustrates the partition of $V \setminus \{v_8\}$, highlighting the vertices in the same partition using a colored or shaded region.

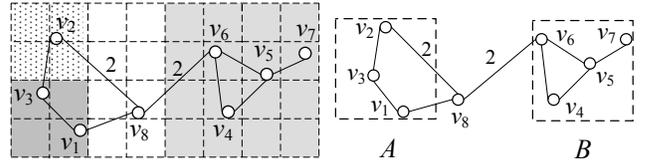

Figure 4: SILC  Figure 5: PCPD

For each vertex $v \in V$, SILC generates a partition of $V \setminus \{v\}$ in the aforementioned manner, and it associates each equivalence class $EC$ with the neighbor of $v$ that lies on the shortest paths from $v$ to the vertices in $EC$. Given any two vertices $s, t \in V$, SILC computes the shortest path between $s$ and $t$ as follows. It first inspects $s$, and examines the partition of $V \setminus \{s\}$ to identify the equivalence class $EC$ that contains $t$. Let $v$ be the neighbor of $s$ that corresponds to $EC$. According to the property of $EC$, the shortest path from $s$ to $t$ must pass by $v$. By inspecting the partition $V \setminus \{v\}$, SILC can identify the neighbor of $v$ that lies on the shortest path from $v$ to $t$. With an iterative application of this traversal method, the complete shortest path from $s$ to $t$ can be obtained.

The above algorithm requires materializing the partition of $V \setminus \{v\}$ corresponding to each vertex $v$. To this end, a straightforward approach is to enumerate the elements of each equivalence class in each partition, which, however, leads to a prohibitive $O(n^2)$ space overhead. To address this issue, Samet et al. [21, 23] propose a concise representation of the partitions, based on the observation that *vertices in the same equivalence class are usually located in the same spatial region*. For example, in the partition of $V \setminus \{v_8\}$ in Figure 4, the vertices in each equivalence class are close to each other, and the three equivalence classes can be covered using three disjoint squares, respectively. Therefore, instead of recording the element of each equivalence class, we may store the square representations of the equivalence classes, so as to save space. In general, for each vertex $v \in V$, the corresponding partition of $V \setminus \{v\}$ can be represented using $O(\sqrt{n})$ disjoint squares [21]. As a consequence, storing the partitions for all vertices in $V$ incurs only an $O(n\sqrt{n})$ space complexity. Furthermore, searching a square region in a partition can be done in $O(\log n)$ time [21]. Hence, SILC can answer any shortest path query in $O(k \log n)$ time, where $k$ is the number of edges in the shortest path. On the other hand, for any distance query between two vertices $s, t \in V$, SILC needs to first compute the shortest path from $s$ and $t$, and then return the sum of the lengths of the edges in the path.

### 3.5 Path-Coherent Pairs Decomposition

Path-Coherent Pairs Decomposition (PCPD) is a technique similar to SILC, in the sense that it also requires pre-computing and compressing all shortest paths among the vertices in the road network. Specifically, PCPD employs a concise representation of shortest paths called *path-coherent pairs*. A path-coherent pair is a triplet $(X, Y, \psi)$, where $X$ and $Y$ are two disjoint square regions, and $\psi$ is either a vertex in $V$ or an edge in $E$, such that $\psi$ lies on the shortest path from any vertex in $X$ to any vertex in $Y$. For instance, Figure 5 illustrates a path-coherent pair $(X, Y, v_8)$ on the road network in Figure 1. Observe that any shortest path from within $X$ to within $Y$ must pass through $v_8$. We say that a path-coherent pair $(X, Y, \psi)$ *covers* two vertices $v_a, v_b \in V$, if and only if $v_a \in X$ and $v_b \in Y$.

Given a road network, PCPD pre-computes a set $S_{pcp}$ of path-coherent pairs, such that any two vertices $v_1, v_2 \in V$ are covered by a *unique* path-coherent pair $(X, Y, \psi) \in S_{pcp}$. With $S_{pcp}$, the shortest path between any two vertices $s, t \in V$ can be com-puted as follows. First, we retrieve the unique path-coherent pair $(X_1, Y_1, \psi_1)$ in $S_{pcp}$ that covers $s$ and $t$. Assume without loss of generality that $\psi_1$ is a vertex in $V$. By the properties of path-coherent pairs, $\psi_1$ should lie on the shortest path from $s$ to $t$. Therefore, we can decompose the shortest path between $s$ and $t$ into two components: the shortest path from $s$ to $\psi_1$ and the shortest path from $\psi_1$ to $t$. After that, we can inspect $S_{pcp}$ to identify a vertex or edge that lies on each of two path components. This enables us to further decompose each component into two smaller parts. By applying the above procedure recursively, we can compute the shortest path from $s$ to $t$ with $O(k)$ lookups in $S_{pcp}$, where $k$ is number of vertices in the shortest path.

| Name | Corresponding Region | Num. of Vertices | Num. of Edges |
|------|---------------------|------------------|---------------|
| DE | Delaware | 48,812 | 120,489 |
| NH | New Hampshire | 115,055 | 264,218 |
| ME | Maine | 187,315 | 422,998 |
| CO | Colorado | 435,666 | 1,057,066 |
| FL | Florida | 1,070,376 | 2,712,798 |
| CA | California and Nevada | 1,890,815 | 4,657,742 |
| E-US | Eastern US | 3,598,623 | 8,778,114 |
| W-US | Western US | 6,262,104 | 15,248,146 |
| C-US | Central US | 14,081,816 | 34,292,496 |
| US | United States | 23,947,347 | 58,333,344 |

**Table 1: Dataset Characteristics**

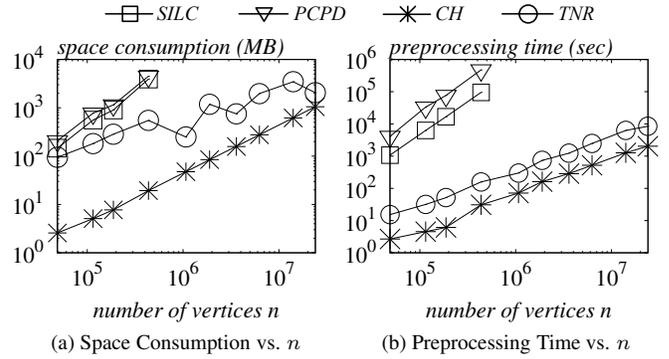

(a) Space Consumption vs. $n$  (b) Preprocessing Time vs. $n$

**Figure 6: Space Overhead and Preprocessing Time vs. $n$**

Sankaranarayanan *et al.* [25] show that each lookup in $S_{pcp}$ can be performed in $O(\log |S_{pcp}|)$ time. Furthermore, under some simplifying assumption about the road network, $|S_{pcp}| = O(n)$ holds. (See Appendix C for a discussion on the simplifying assumption.) This indicates that PCPD has an $O(n)$ space complexity, and its time complexity for shortest path queries is $O(k \log n)$. For any distance query between two vertices $s, t \in V$, PCPD first computes the shortest path between $s$ and $t$, and then returns the length of the path.

## 4. EXPERIMENTS

This section evaluates the performance of the five techniques introduced in Section 3, namely, the bidirectional Dijkstra's algorithm [19], CH [11], TNR [5], SILC [21, 23], and PCPD [25], in terms of their space overhead, preprocessing time, and query efficiency (for both shortest path and distance queries).

### 4.1 Experimental Settings

We adopted the implementation of CH from its inventors [1], and we implemented TNR, SILC, and PCPD from scratch as the source codes of those three methods were unavailable. (See Appendix D for implementation details.) The source code of our implementation is available for download [2]. All the algorithms were coded using Microsoft's Visual C++ 2008, and they used common subroutines for similar tasks. We conducted experiments on a computer running Windows 7 with an Intel Xeon 2.67 GHz CPU and 24 GB RAM.

As discussed in Section 3, the preprocessing steps of SILC and PCPD require computing all-pairs shortest paths in a road network, while TNR requires identifying the shortest paths from within each grid cell to beyond the outer shell of the cell. For efficiency, we employed CH to accelerate the shortest path computation required in the preprocessing steps of SILC, PCPD, and TNR. In addition, recall that the performance of TNR is affected by (i) the granularity of the grid imposed on the road network and (ii) the alternative technique that is used to handle queries that TNR cannot answer. We tested various grid granularities and we considered two methods, CH and the bidirectional Dijkstra's algorithm, for handling the shortest path and distance queries that cannot be processed by TNR. We report the results when TNR is combined with CH on a $128 \times 128$ grid, since this combination incurs a smaller space overhead and achieves higher query efficiency than all of other alternatives do (see Appendix E.1 for a detailed comparison).

We consider that the indexing structures of all techniques should be memory resident, so as to ensure responsive query processing (which is crucial to online map services and commercial navigation systems). Accordingly, we report the results of a technique on a dataset only when the size of its indexing structure is less than 24 GB, *i.e.*, the size of the main memory of our computer.

### 4.2 Datasets and Queries

We used ten datasets of various sizes from the Ninth DIMACS Implementation Challenge [3]. Each dataset contains an undirected graph that represents a part of the road network in the United States. Each edge in a graph represents the time required to travel between the two endpoints of the edge. Table 1 illustrates the numbers of vertices and edges in the data.

On each dataset, we generated ten sets $Q_1, Q_2, \ldots, Q_{10}$ of queries as follows. We first imposed a $1024 \times 1024$ grid on the road network and computed the side length $l$ of each grid cell. After that, we randomly selected ten thousand pairs of vertices from the road network to compose $Q_i$ ($i \in [1, 10]$), such that the $L_\infty$ distance between each pair of vertices is in $[2^{i-1} \cdot l, 2^i \cdot l)$. Note that the $L_\infty$ distance between two vertices of each query in $Q_i$ is larger than that in $Q_{i-1}$. For each query set $Q_i$ ($i \in [1, 10]$) and each technique, we report the average running time of the technique over all queries in the set. We have also tested another ten sets of queries generated based on the road network distances among the vertices, and we include the experimental results on those query sets in Appendix E.2.

### 4.3 Space Overhead and Preprocessing Time

In the first set of experiments, we measured the pre-computation time of CH, TNR, SILC, and PCPD, as well as the size of the index structure generated by each technique. Figure 6(a) shows the space overhead of each technique as a function of $n$, *i.e.*, the number of vertices in each dataset. The overhead of CH is the smallest among all the techniques, and is roughly linear to $n$. The space consumption of TNR is significantly higher than that of CH when $n$ is small, but the disparity becomes smaller as $n$ increases. To understand this, recall that TNR pre-computes two sets of distance information: (i) the set $I_1$ of distances between any pair of access nodes, and (ii) the set $I_2$ of distances from any vertex $v$ to any access node of the grid cell that contains $v$. We observed in our experiment that the number of access nodes for each grid cell is

410

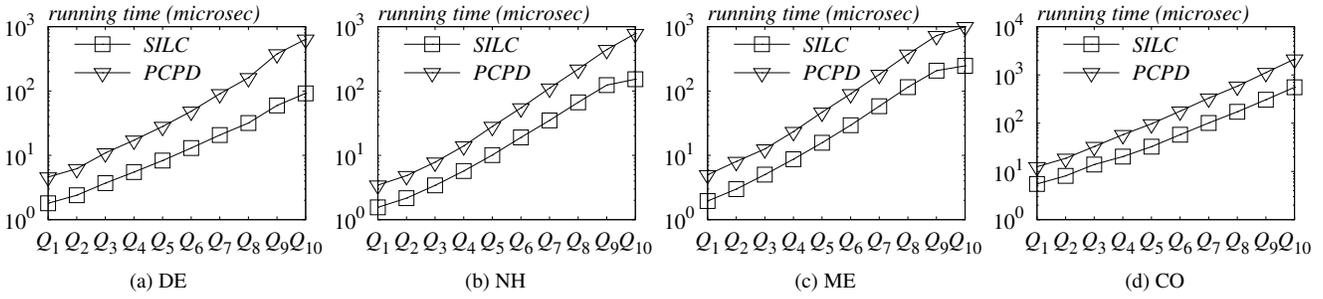

**Figure 7: SILC vs. PCPD on Shortest Path Queries**

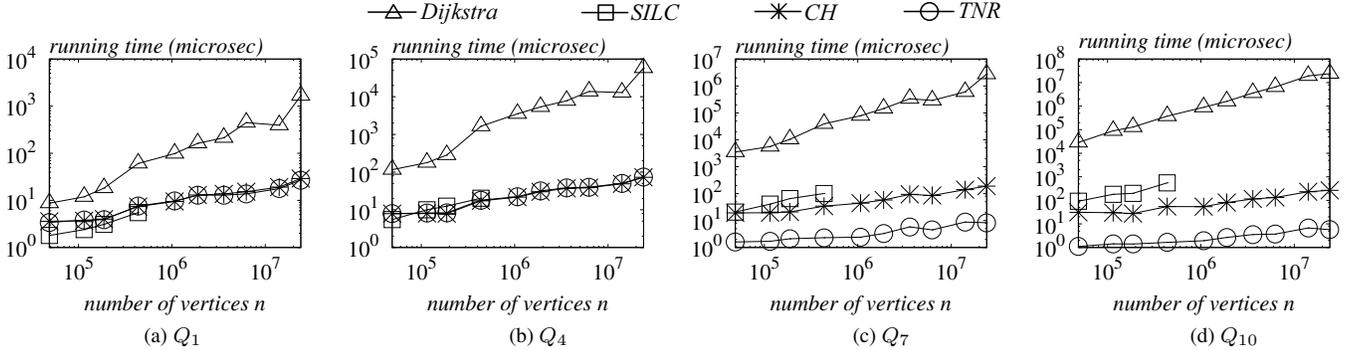

**Figure 8: Efficiency of Distance Queries vs. $n$**

roughly a constant (around 10) on all the road networks. Therefore, the space requirement of $I_1$ is similar for all datasets. On the other hand, $I_2$ takes $O(n)$ space, and it incurs a comparable overhead with CH. On small datasets, $I_1$ dominates the space overhead of TNR, which accounts for the significant performance gap between TNR and CH. On large road networks, however, $I_2$ becomes the major contributing factor of TNR's space requirement. Therefore, the gap of space consumption becomes smaller.

On the other hand, SILC and PCPD entail space overheads that are orders of magnitude higher than that of CH. For example, on the CO dataset with $0.4$ million vertices, the index structures constructed by SILC and PCPD are both over $4$ GB in size. Furthermore, the overheads of SILC and PCPD are similar, even though the space complexities of SILC and PCPD are $O(n \log n)$ and $O(n)$, respectively. This indicates that the complexity of PCPD comes with a large hidden constant (see Appendix C for a discussion). We only report the cost of SILC and PCPD on the four smallest datasets; for datasets with more than one million vertices, the space consumptions of SILC and PCPD exceed $24$ GB.

Figure 6(b) illustrates the preprocessing time of each technique. CH incurs the smallest pre-computation overhead among the four methods. For example, it requires only 30 minutes to process the US dataset, which contains more than 20 million vertices. The preprocessing cost of TNR is consistently higher than that of CH, since the computation of access nodes involves a large number of shortest path queries, which entail considerable overhead. The preprocessing time of SILC and PCPD is orders of magnitude higher than that of CH, since both techniques require pre-computing the all-pairs shortest paths in a road network. For example, on the CO dataset with $0.4$ million vertices, the pre-computation time of SILC (resp. PCPD) is over 26 (resp. 132) hours.

### 4.4 SILC vs. PCPD

The second set of experiments compared SILC with PCPD on the efficiency for shortest path queries. Figure 5 illustrates the average running time of SILC and PCPD in processing the shortest path queries on the four smallest datasets. Regardless of the query set and dataset, SILC consistently outperforms PCPD. To explain this, recall that both SILC and PCPD answer a shortest path query using $k$ lookups in their respective indexing structures, where $k$ is the number of edges in the shortest path. In particular, SILC performs each lookup over a set of $O(\sqrt{n})$ disjoint square regions, and it returns the region that contains the destination of the shortest path. On the other hand, PCPD conducts each lookup over a set of $O(n)$ path-coherent pairs, and it identifies the pair that covers both the source and destination of the shortest path. With proper index structures, each lookup of SILC (resp. PCPD) can be performed by inspecting $O(\log n)$ square regions (path-coherent pairs). Nevertheless, checking whether a square region contains a vertex is more efficient than deciding whether a path-coherent pair covers a pair of vertices. Hence, SILC incurs less query overhead than PCPD does. For distance queries, the relative superiority of SILC and PCPD remain the same as in Figure 5, since both SILC and PCPD answer a distance query by first computing the corresponding shortest path and then returning the length of the path.

To sum up, our results show that SILC outperforms PCPD in terms of query efficiency and preprocessing time, and the space overhead of SILC is similar to that of PCPD (see Figure 6). These findings complement the existing work [25], which establishes the superiority of PCPD over SILC *in terms of asymptotic space complexity*, but does not compare the practical performance of PCPD and SILC. Since PCPD is dominated by SILC in almost all aspects, we omit the results for PCPD in the following sections.

### 4.5 Efficiency of Distance Queries

The next set of experiments evaluated the performance of each technique for distance queries. Figure 8 shows the average running time of each technique as a function of $n$. The bidirectional Dijkstra's algorithm is much slower than the other three methods. On $Q_1$, SILC performs slightly better than the others on the four datasets with fewer than one million nodes. On $Q_4$, however, the performance of SILC, CH, and TNR becomes similar. Meanwhile, TNR performs an order of magnitude better than CH and SILC on $Q_7$ and $Q_{10}$. Figure 9 illustrates the running time of SILC, CH, and TNR as the query set varies. Observe that the query time of SILC increases considerably with the $L_\infty$ distance between the two vertices $s$ and $t$ in the query. This is due to the fact that SILC takes $O(k \log n)$ time to answer each query, where $k$ is the number of edges in the shortest path between $s$ and $t$. When the $L_\infty$ distance between $s$ and $t$ increases, $k$ tends to be larger, leading to a higher processing cost for SILC. In contrast, the running time of CH does not increase significantly with the $L_\infty$ distance between $s$ and $t$, since CH can utilize the shortcuts to identify $dist(s,t)$ without traversing all edges in the shortest path from $s$ to $t$. As a consequence, CH outperforms SILC when $s$ and $t$ are far apart.

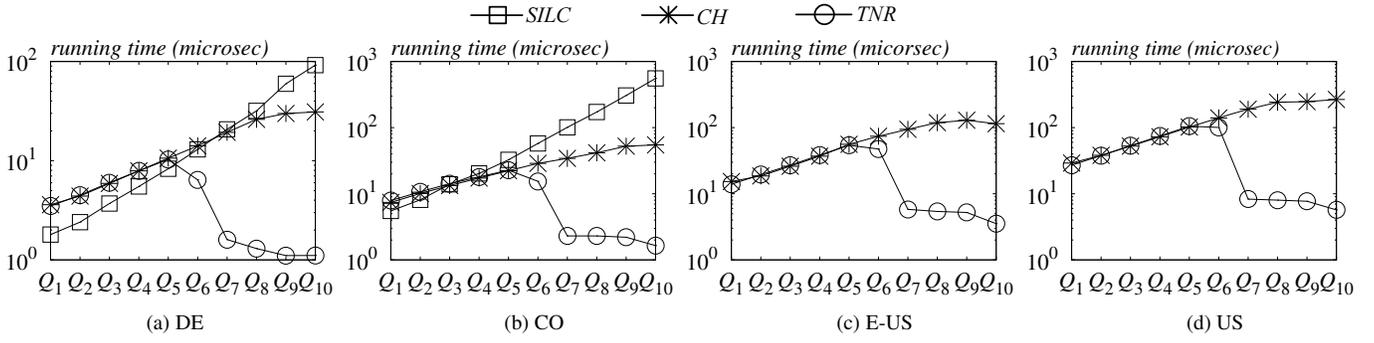

Figure 9: Efficiency of Distance Queries vs. Query Sets

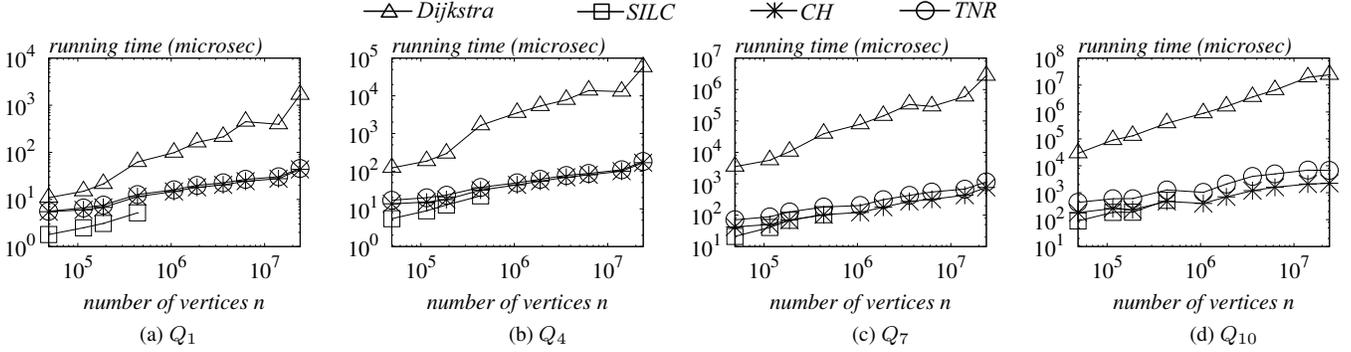

Figure 10: Efficiency of Shortest Path Queries vs. $n$

Notice that TNR considerably outperforms CH on query sets $Q_7, Q_8, Q_9, Q_{10}$. This is because, for each query in those four sets, TNR can answer the query by inspecting a few pre-computed distances; in contrast, CH needs to invoke the bidirectional Dijkstra's algorithm to process the query, which leads to a higher computation overhead. On the other hand, in query sets $Q_1, Q_2, \ldots, Q_5$, each query corresponds to two vertices that are close to each other; TNR cannot process such queries based only on the pre-computed information, and it needs to resort to an alternative technique (*i.e.*, CH). Hence, TNR and CH perform identically on $Q_1, Q_2, \ldots, Q_5$. For $Q_6$, TNR can handle part of the queries without applying CH, and therefore, the time required in processing $Q_6$ is shorter than that demanded by $Q_5$.

### 4.6 Efficiency of Shortest Path Queries

The last set of experiments investigated the performance of each technique for shortest path queries. Figure 10 plots the running time of each technique as a function of $n$. Again, the query time of the bidirectional Dijkstra's algorithm is orders of magnitude higher than that of the other techniques. Meanwhile, SILC outperforms CH on the four smallest datasets, for which the indexing structure of SILC does not exceed 24 GB. The running time of SILC is identical with that for distance queries (see Figure 8), since SILC employs the same algorithm to answer both shortest path and distance queries. In contrast, the query overhead of CH is considerably higher than the case for distance queries. Recall that, for any distance query between two vertices $s$ and $t$, CH answers the query by computing the *augmented* shortest path between $s$ and $t$, *i.e.*, a path that may consist of shortcuts. To identify the exact shortest path between $s$ and $t$, CH would need to transform the augmented shortest path into a path that contains only the edges in $E$, which incurs extra overhead. As a consequence, CH is less efficient for shortest path queries (than for distance queries).

We also observed that TNR performs no better than CH in all cases. Specifically, the running time of TNR and CH is identical on $Q_1, \ldots, Q_5$, since (i) each query in these sets corresponds to two vertices that are close to each other, and (ii) TNR needs to resort to CH to answer such queries. On $Q_7, \ldots, Q_{10}$, however, TNR entails a higher overhead than CH does. This is because TNR answers each shortest path query by invoking $O(k)$ distance queries, where $k$ denotes the number of edges in the shortest path. As the $L_\infty$ distance between the source and destination of each query becomes larger (from $Q_7$ to $Q_{10}$), the number $k$ of edges in each shortest path tends to increase, and thus, TNR needs to invoke more distance queries. This explains why the disparity between CH and TNR becomes larger from $Q_7$ to $Q_{10}$. Finally, since TNR can answer only part of the queries in $Q_6$, its overhead on $Q_6$ is only slightly higher than that of CH.

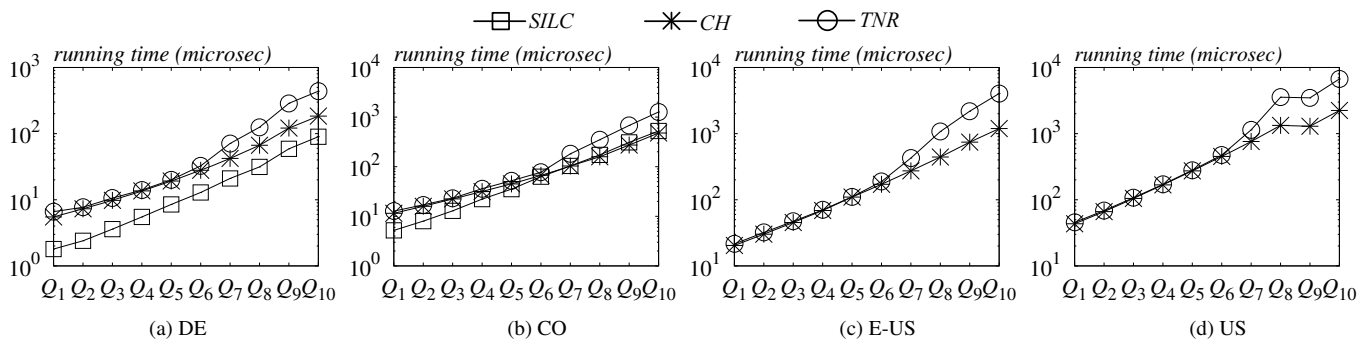

Figure 11: Efficiency of Shortest Path Queries vs. Query Sets

## 4.7 Summary of Experimental Results

As a summary, we have the following observations from our experimental results. First, both SILC and PCPD incur significant preprocessing time and space overhead, which renders them inapplicable for large road networks with millions of vertices. Second, PCPD is inferior to SILC in terms of pre-computation cost, space consumption, as well as query efficiency. Third, SILC outperforms both CH and TNR on shortest path queries, but it is not as efficient as CH and TNR on distance queries. Fourth, CH entails minimum preprocessing and space cost, and yet, it offers excellent performance for both shortest path and distance queries. Finally, TNR significantly improves over CH on distance queries, but the improvement comes at the cost of considerable pre-computation and space overhead.

## 5. CONCLUSIONS

This paper presents an experimental comparison of four state-of-the-art techniques for answering shortest path and distance queries on road networks, namely, SILC, PCPD, CH, and TNR. We used a variety of real datasets with up to twenty million vertices, and we evaluated each technique in terms of its preprocessing time, space overhead, and query efficiency. From our experimental results, we have the following observations. First, CH is the most space-economic technique compared with TNR, SILC and PCPD, and yet, it is the second most efficient technique in answering shortest path and distance queries. This makes CH a preferable choice when both space efficiency and time efficiency are major concerns. Second, TNR can be combined with CH to achieve significant speedup for distance queries, especially when the source and destination vertices are far away from each other. However, it also entails considerable space overhead, and it is not as efficient as CH for shortest path queries. Third, SILC incurs significant preprocessing time and space consumption, but it offers superior efficiency for shortest path queries. Therefore, SILC is recommended for processing shortest path queries when time efficiency is crucial and space overhead is less concerned. Finally, although PCPD was proposed as a successor to SILC with an improved asymptotic space complexity, its practical performance (in terms of preprocessing time, space consumption, and query efficiency) is inferior to SILC. As a consequence, PCPD is an unfavorable choice (compared to SILC) for indexing road networks.

## 6. ACKNOWLEDGMENTS

This work was supported by the Nanyang Technological University under SUG Grant M58020016 and AcRF Tier 1 Grant RG 35/09. The authors would like to thank R. Geisberger and P. Sanders for providing their implementation of Contraction Hierarchies (CH).

# APPENDIX
## A. ADDITIONAL RELATED WORK

Apart from the techniques evaluated in this paper, many other solutions have been proposed for shortest path and distance queries. Representative methods among them include *ALT* [12], *RE* [13], *Arc Flags* [15], *Highway Hierarchies* [22], *HiTi* [17], and *HEPV* [16]. In particular, ALT preprocesses the road network by first selecting a small set of vertices, called the *landmarks*. Then, it pre-computes the distance from each vertex in $V$ to each landmark. With the pre-computed distances, we can efficiently derive a lower-bound of $dist(s, v) + dist(v, t)$ for any three vertices $s$, $v$, and $t$. ALT incorporates such lowerbounds with Dijkstra's algorithm to improve query efficiency.

Similar to ALT, RE [13] also pre-computes certain information about each vertex $v$ (referred to as the *reach* of $v$) to accelerate query processing. Specifically, for any shortest path that passes through $v$, the reach of $v$ is an upperbound on $\min\{dist(s', v), dist(v, t')\}$, where $s'$ and $t'$ are the source and destination of the path, respectively. Observe that, given any two vertices $s$ and $t$, if the reach of $v$ is smaller than both $dist(s, v)$ and $dist(v, t)$, then $v$ cannot be on the shortest path from $s$ to $t$. Based on this observation, RE incorporates the reach of each vertex with the bidirectional Dijkstra's algorithm to accelerate shortest path and distance queries.

Arc Flags [15] is a method similar to SILC in the sense that it also imposes a grid on the road network. In the preprocessing step, for each vertex $v$ and each edge $e$ incident to $v$, Arc Flags tags $e$ with the grid cells in which there is at least one vertex $v'$ whose shortest path to $v'$ passes through $e$. Then, given any two vertices $s$ and $t$, Arc Flags can efficiently identify the shortest path or distance between $s$ and $t$ by applying a revised version of Dijkstra's algorithm that avoids visiting irrelevant edges. Highway Hierarchies [22] is a predecessor to CH that constructs a partial order on the vertices. It organizes the vertices in the road network into a hierarchy based on their relative importance, and it creates shortcuts among vertices at the same level of the hierarchy to improve query efficiency.

HiTi [17] is a technique that pre-computes a set of vertex-disjoint partitions of the road network. For each component $C$ of a partition, it pre-computes the distance between any two *boundary* vertices of $C$, i.e., vertices in $P$ that are adjacent to the vertices in other partitions. With these pre-computed distances, HiTi employs a modified version of Dijkstra's algorithm for efficient processing. The algorithm, however, is only applicable when the weight on each edge of road network represents the Euclidean distance between the two endpoints of the edge. This considerably restricts the application of HiTi in practice. In particular, HiTi cannot handle the datasets used in our experiments, since each of our datasets contains edges whose weights do not represent the Euclidean distances among endpoints; instead, the weight of each edge represents the time required to traverse the edge. HEPV [16] is a predecessor of HiTi that also pre-processes the road network by partitioning the graph and pre-computing the distances among certain vertices in each partition component. Compared with HiTi, the major deficiency of HEPV is that it incurs a huge space consumption, which renders it unsuitable for medium and large road networks [17].

All of the aforementioned methods, except HiTi and HEPV, are previously shown to be inferior to CH in terms of both space overhead and query performance [26]. Delling *et al.* [8] demonstrate that Arc Flags can be combined with CH to further reduce query overhead, while Bast *et al.* [6] present a hybrid method that incorporates Highway Hierarchies and TNR. We do not consider those combinations in our paper, since they incur significantly higher pre-processing overhead or space consumption, and they are considerably more complicated to implement (which renders them less likely to be adopted in practice).

In addition, there exist several theoretical studies (*e.g.*, [4, 10, 14, 18]) on shortest path and distance queries. Furthermore, there are also a few variations of the techniques evaluated in this paper. In particular, Sankaranarayanan and Samet [24] propose a revised version of PCPD that can handle *approximate* distance queries efficiently; Samet et al. [21] show that SILC can also be used to achieve superior performance for *nearest neighbor queries*; Rice and Tsotras [20] extend CH for answering shortest path and distance queries with restrictions on the properties of the edges that comprise the shortest path.

## B. DEFECTS OF TNR

Recall that TNR [5] requires pre-computing the access nodes of each cell $C$ in a grid imposed on the road network. As discussed in Section 3.3, a simple solution for this pre-computation is to first (i) derive the shortest paths from each vertex in $C$ to the vertices on the boundary of $C$'s outer shell, and then (ii) select appropriate vertices on the shortest paths to form a set of access nodes. This solution, when implemented with Dijkstra's algorithm, incurs considerable overhead [5]. The reason is that the solution requires applying Dijkstra's algorithm $O(n)$ times and, in each time, the algorithm needs to search over roughly $9 \times 9$ cells in the grid (*i.e.*, the number of cells contained in the outer shell of each grid cell). For better efficiency, Bast et al. [5] propose an improved approach that also applies Dijstra's algorithm $O(n)$ times, but it significantly reduces the search space of the algorithm. In the following, we will explain Bast et al.'s approach using the example in Figure 12, which shows a grid cell $C_0$ as well as its inner and outer shells.

Table 2: Upperbound of $\delta$ in Road Networks

| Dataset | DE | NH | ME | CO | FL | CA | E-US | W-US | C-US | US |
|---|---|---|---|---|---|---|---|---|---|---|
| $\min\{length(P')/length(P)\}$ | 1 | 1.00001 | 1.00127 | 1.00029 | 1.00003 | 1.00379 | 1 | 1.0001 | 1.00104 | 1.00046 |

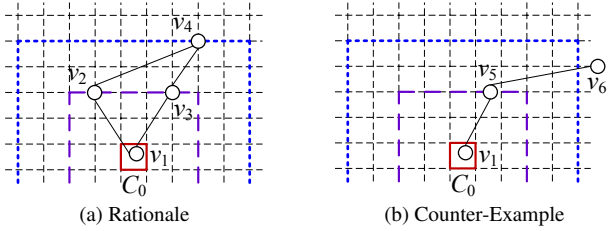

(a) Rationale  (b) Counter-Example

**Figure 12: Illustration of Bast et al.'s Approach**

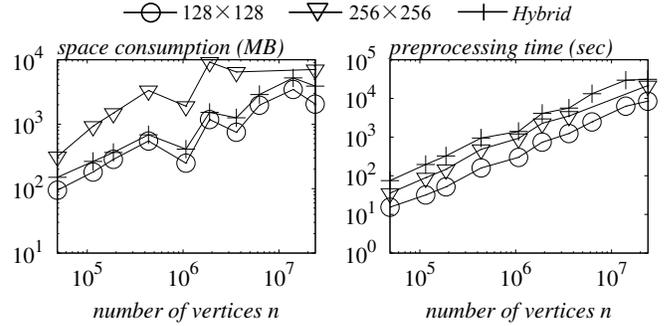

(a) Space Consumption vs. $n$   (b) Preprocessing Time vs. $n$

**Figure 13: Space Overhead and Preprocessing Time vs. $n$**

Let $S_{in}$ ($S_{up}$) be the set of vertices in $V$ that are adjacent to the edges intersecting the upper boundary of $C_0$'s inner (outer) shell. Bast et al. claim that, a vertex $v \in S_{in}$ can be an access node for $C_0$, if and only if $v$ is on the shortest path from a vertex in $C_0$ to a vertex in $S_{up}$. Based on this claim, Bast et al. propose to identify the access nodes in $S_{in}$ as follows. First, for each vertex $v_j \in S_{in}$, we apply Dijkstra's algorithm to compute the distance from $v_j$ to each vertex in $C_0$ and $S_{up}$. (Notice that the search space for Dijkstra's algorithm is much smaller than $9 \times 9$ cells in this case.) Then, for each vertex $v_i$ in $C_0$ and each vertex $v_k \in S_{up}$, we identify the vertex $v \in S_{in}$ that minimizes $dist(v_i, v_j) + dist(v_j, v_k)$, and we mark $v$ as an access node for $C_0$. For example, assume that $S_{in}$ contains only two vertices $v_2$ and $v_3$, as shown in Figure 12(a). For the vertex $v_1$ in $C_0$ and the vertex $v_4 \in S_{up}$, we have $dist(v_1, v_2) + dist(v_2, v_4) < dist(v_1, v_3) + dist(v_3, v_4)$. Therefore, $v_2$ will be tagged as an access node for $C_0$. After all vertices in $C_0$ and $S_{up}$ are examined, we obtain the access nodes for $C_0$ that are adjacent to the edges intersecting the upper boundary of $C_0$'s inner shell. With a similar approach, we can also derive the access nodes for $C_0$ that are pertinent to the edges intersecting the other three sides of the inner shell, which leads to a "complete" set of access nodes for $C_0$. Bast et al. show that, with proper optimization, the aforementioned approach can compute the access nodes for all cells by invoking Dijkstra's algorithm at most once for each vertex in $V$.

Nevertheless, Bast et al.'s approach is flawed as it is developed based on an invalid claim. Specifically, even if a vertex $v \in S_{in}$ is an access node for $C_0$, it is not necessarily on the shortest path from some vertex in $C_0$ to a vertex in $S_{up}$. To illustrate this, consider the example in Figure 12(b), where $S_{in}$ contains a vertex $v_5$ that has only two neighbors, namely, a vertex $v_1$ in $C_0$ and a vertex $v_6$ in the exterior of $C_0$'s outer shell. Assume that $v_5$ is $v_6$'s only neighbor. Then, $v_5$ must be an access node for $C_0$, since it is the only vertex that connects $v_1$ to $v_6$. However, Bast et al.'s approach would omit $v_5$, since $v_5$ is not on the shortest path from any vertex in $C_0$ to any vertex in $S_{up}$. As a consequence, the set of access nodes computed by Bast et al.'s approach is incomplete, since there exists a shortest path from within $C_0$ to beyond $C_0$'s outer shell (i.e., the path from $v_1$ to $v_6$ via $v_5$) that is not covered by any access node. The incomplete set of access nodes, when used to answer shortest path and distance queries, could lead to incorrect results.

To remedy the defect of Bast et al.'s approach, we resort to the simple solution that identifies the access nodes for each cell $C$, by computing the shortest paths from each vertex in $C$ to the vertices at the boundary of $C$'s outer shell. Instead of directly applying Dijkstra's algorithm to derive the shortest paths, however, we construct contraction hierarchies on the road network in advance, so as to reduce the computation cost of deriving access nodes. Our experiments show that the pre-computation overhead incurred by constructing contraction hierarchies is negligible compared with the reduction in the cost of access node computation.

## C. SPACE COMPLEXITY OF PCPD

The space complexity of PCPD was proved under the assumption that the shortest path between any two vertices $s$ and $t$ in the road network is $\delta$-*redundant* [25]. Specifically, the shortest path $P$ between $s$ and $t$ is $\delta$-redundant ($\delta > 1$), if and only if for any path $P'$ from $s$ to $t$ that does not share any common vertex with $P$, the length of $P'$ is at least $\delta$ times the length of $P$. The path $P'$ is referred to as a *core-disjoint path* [25] between $s$ and $t$. Sankaranarayanan et al. [25] show that, on any $\delta$-redundant road network, the space complexity of PCPD is $O((2 + \frac{2}{\delta-1})^2 \cdot n)$, which is linear to $n$ when $\delta$ is regarded as a constant.

As acknowledged by Sankaranarayanan et al. [25], however, real road networks may not always be $\delta$-redundant. Furthermore, even if a road network is $\delta$-redundant, the value of $\delta$ might be rather close to 1, which results in a enormous constant factor (*i.e.*, $(2 + \frac{2}{\delta-1})^2$) in the space complexity of PCPD. To demonstrate this, we conducted an experiment as follows. For each of the ten datasets in Table 1 and for each pair of vertices $(s, t)$ in our query sets $Q_1, Q_2, \ldots, Q_{10}$, we computed the shortest path $P$ and shortest core-disjoint path $P'$ between $s$ and $t$, and we calculated the length of $P'$ as a fraction of the length of $P$, denoted as $length(P')/length(P)$. Table 2 shows the minimum value of $length(P')/length(P)$ (which is an upperbound of $\delta$) that we observed on each dataset. The values are either equal or very close to 1, which explains why PCPD incurs significant space overhead on our datasets (see Section 4.3).

## D. IMPLEMENTATIONS

We implemented TNR, SILC and PCPD based on the graph data structure used in the source code of CH provided by its inventors [1]. The structure is essentially a hash table consisting of two arrays. Each element in the first array stores the information of a vertex $u$, as well as a pointer to a block of elements in the second array, such that each element in the block corresponds to an edge adjacent to $u$. In other words, for any edge that connects two vertices $u$ and $v$, it is stored repeatedly in the blocks that correspond to $u$ and $v$, respectively.

For TNR, we employ the algorithm outlined in the end of Section 3.3 to compute the access nodes for each cell. Then, we use three hash tables to store the information related to the access nodes. The first table associates any grid cell $C$ to the set of access nodes for $C$. The second table records the distance from each vertex $v$ to each access node $a$ for the cell that contains $v$. That is, the table maps the ordered pair $(v, a)$ to $dist(v, a)$. The third table maintains the distance between any pair of access nodes.

For SILC, we apply Dijkstra's algorithm to derive the partition of $V \setminus \{v\}$ for any vertex $v$. After that, the concise representation of each partition is computed using the method suggested by Samet et al. [21]. We first impose a $2 \times 2$ grid on the road network, and we inspect the vertices contained in each grid cell $C$. If there exist two vertices in $C$ that are from two different equivalence classes, $C$ is further divided into four quadrants. This procedure is recursively applied on the grid cells, until each cell contains only vertices from the same equivalence class. After that, each cell is transformed into an interval on a two-dimensional Z-curve [21], and the intervals corresponding to the same partition are stored in a binary search tree to accelerate query processing.

For PCPD, we compute the path-coherent pairs of the road network using the following approach proposed by Sankaranarayanan et al. [25]. First, we construct a pair of square regions $(X, Y)$, such that both $X$ and $Y$ cover all vertices in $V$. After that, we compute the shortest path from any vertex in $X$ to any vertex in $Y$. If all shortest paths share a common vertex or edge, we construct a path-coherent pair $(X, Y, \psi)$, where $\psi$ denotes the vertex or edge shared by the shortest paths. Otherwise, we divide $X$ (resp. $Y$) into four quadrants $X_1, X_2, X_3, X_4$ (resp. $Y_1, Y_2, Y_3, Y_4$), and we replace $(X, Y)$ with 16 pairs of square regions $(X_i, Y_j)$ $(i, j \in [1, 4])$. The aforementioned procedure is recursively applied on each pair of square regions, until all pairs of squares are transformed into path-coherent pairs. To reduce the cost of testing whether all shortest paths from $X_i$ to $Y_j$ pass by a common vertex or edge, we implement the test as a nested loop over the vertices in $X_i$ and $Y_j$, and we maintain the set of vertices and edges shared by the shortest paths that we have examined. Once the set becomes empty, we declare that $X_i$ and $Y_j$ cannot form a path-coherent pair, and we proceed to further divide $X_i$ and $Y_j$.

## E. ADDITIONAL EXPERIMENTS

### E.1 Alternative Implementations of TNR

This section evaluates the performance of TNR when the grid granularity varies, and when different techniques are adopted for processing the queries that cannot be handled by TNR. The first set of experiments investigates the preprocessing and space overhead of TNR using a $128 \times 128$ grid $D_{128}$, a $256 \times 256$ grid $D_{256}$, and a hybrid grid that combines $D_{128}$ and $D_{256}$ [5]. Specifically, when the hybrid grid is adopted, we first calculate all access nodes in $D_{128}$ and $D_{256}$; then, we pre-compute the distances between all (resp. some) pairs of access nodes in $D_{128}$ (resp. $D_{256}$). In particular, given an access node $a_1$ for a cell $C_1$ and another access node $a_2$ for a cell $C_2$ in $D_{256}$, we pre-compute $dist(a_1, a_2)$ only when the outer shells of $C_1$ and $C_2$ overlap with each other. This is because, when the outer shells of $C_1$ and $C_2$ are disjoint, the distance between any vertex in $C_1$ and any vertex in $C_2$ can be derived using the access nodes on $D_{128}$, which renders it redundant to pre-compute $dist(a_1, a_2)$.

Figure 13(a) illustrates the space overhead of TNR as a function of $n$. The overhead of $D_{128}$ is smaller than that of the hybrid grid, since the distance information pre-computed by the former is a strict subset of the information preprocessed by the latter. In turn, the hybrid grid consumes less space than $D_{256}$, since $D_{256}$ requires storing the pairwise distances among all access nodes on a $256 \times 256$ grid, while the hybrid grid records only a small subset of those distances. Figure 13(b) shows the preprocessing time of TNR when $n$ varies. $D_{256}$ entails a higher pre-computation cost than $D_{128}$ does, as it requires deriving a larger set of access nodes. The overhead of the hybrid grid is the largest, since it needs to process all access nodes in both $D_{128}$ and $D_{256}$.

Note that the space required by $D_{256}$ exceeds 24 GB on the W-US and C-US datasets, and hence, Figure 13 does not show the results of $D_{256}$ on those road networks. Compared with the hybrid grid, $D_{256}$ incurs a significantly higher space overhead, and yet, it does not enable TNR to answer more (shortest path or distance) queries without invoking an alternative technique. Therefore, we will omit $D_{256}$ in the following experiments.

The next set of experiments evaluates the efficiency of TNR when it adopts CH and the bidirectional Dijkstra's algorithm for processing the queries that cannot be handled by TNR. Figure 14 illustrates the running time of TNR for distance queries. Regardless of whether $D_{128}$ or the hybrid grid is used, TNR performs significantly better when it is incorporated with CH instead of the bidirectional Dijkstra's algorithm. This justifies the (small) additional space overhead incurred by combining CH with TNR. Furthermore, when the bidirectional Dijkstra's algorithm is adopted, $D_{128}$ and the hybrid grid result in almost identical query performance, except that the running time on the hybrid grid is slightly lower on the query sets $Q_5$ and $Q_6$. This is because, TNR answers any query in $Q_1, \ldots, Q_4$ (resp. $Q_7, \ldots, Q_{10}$) by invoking CH (resp. inspecting the access nodes on the $128 \times 128$ grid), regardless of whether $D_{128}$ or the hybrid grid is applied. On the other hand, there exist queries in $Q_5$ and $Q_6$ that can be handled with the hybrid grid but not $D_{128}$, which leads to the discrepancy between the two approaches on $Q_5$ and $Q_6$. A similar phenomenon can be observed when CH is incorporated: $D_{128}$ entails smaller overhead than the hybrid grid on $Q_6$, which indicates that CH is more efficient in answering queries in $Q_6$ than the hybrid grid does. Figure 15 illustrates the computation time of TNR for shortest path queries. The results are qualitatively similar to those in Figure 14.

In summary, TNR performs the most efficiently when it employs a $128 \times 128$ grid and applies CH as the alternative technique for query answering. Therefore, we adopts this setting of TNR in our experiments.

### E.2 Alternative Query Sets

This section investigates the query performance of the bidirectional Dijkstra's algorithm, SILC, CH, and TNR on ten alternative sets of queries $R_1, \ldots, R_{10}$. These query sets were generated based on the road network distances among the vertices, which contrast the query sets in Sections 4.5 and 4.6 as they were generated based on the $L_\infty$ distances. Specifically, on each dataset, we first computed a rough estimation of the maximum distance $l_d$ between any two vertices. After that, we inserted 10000 pairs of vertices $(u, v)$ into $R_i$ ($i \in [1, 10]$), such that $dist(u, v) \in [2^{i-11} \cdot l_d, 2^{i-10} \cdot l_d)$. Note that the length of the shortest path between any pair of vertices in $R_i$ is larger than that in $R_{i-1}$.

Figure 16 (resp. 17 ) shows the running time of each technique for the distance queries (resp. shortest path queries) in $R_i$ ($i \in [1, 10]$). The results are qualitatively similar to those presented in Figures 8 and 10, which confirms our findings in Section 4 about the relative superiority of each technique.

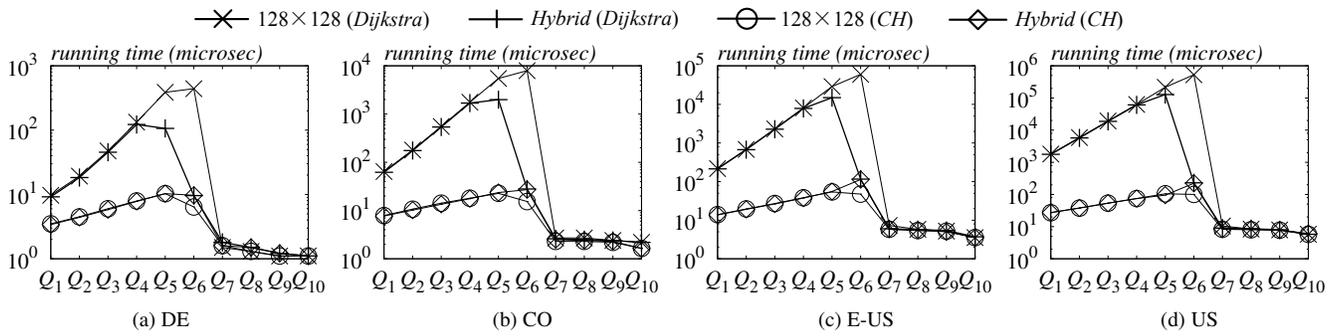

**Figure 14: Efficiency of Distance Queries vs. Query Sets**

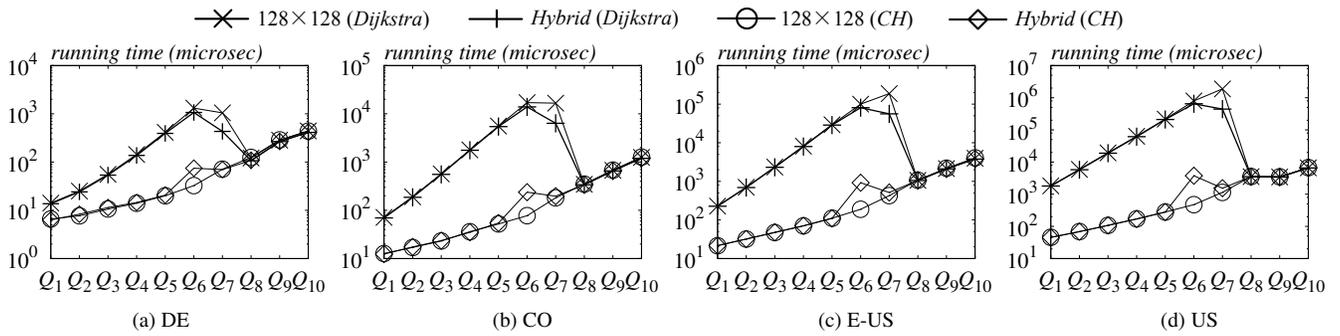

**Figure 15: Efficiency of Shortest Path Queries vs. Query Sets**

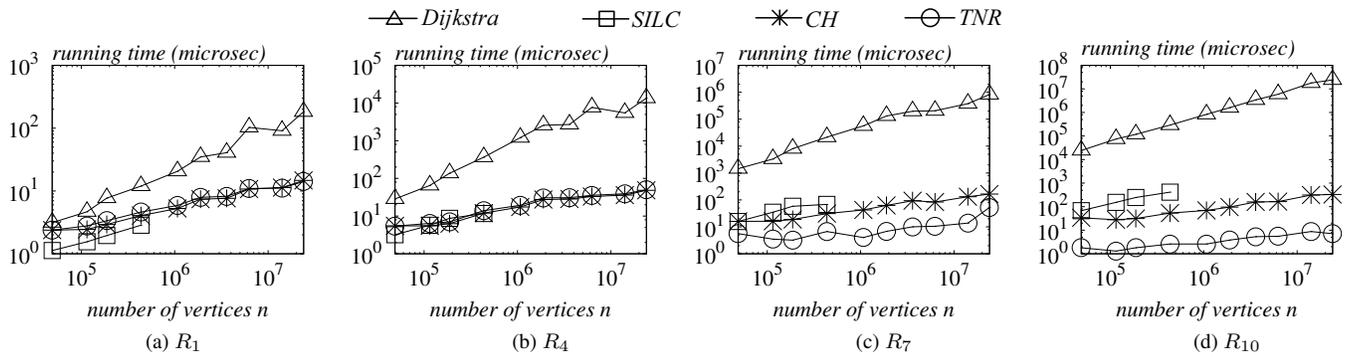

**Figure 16: Efficiency of Distance Queries vs. $n$**

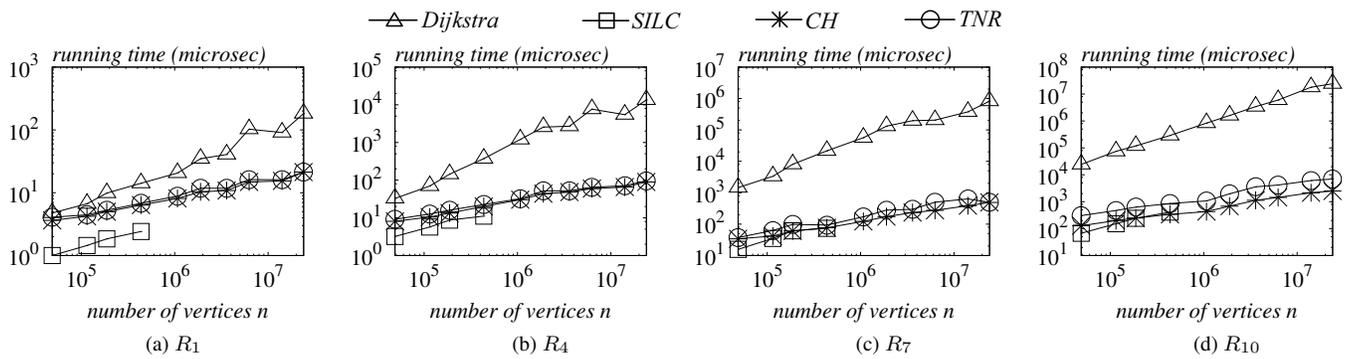

**Figure 17: Efficiency of Shortest Path Queries vs. $n$**